\documentclass[a4paper,12pt]{amsart}
\usepackage{amsfonts,amsmath,amsthm}
\usepackage{amsmath,amssymb}  
\usepackage{graphicx}

\newcommand*{\scrH}{\mathord{\mathcal{H}}}%
\newcommand*{\scrN}{\mathord{\mathcal{N}}}%
\newcommand*{\AAA}{{\mathbb{A}}}
\newcommand*{\NN}{{\mathbb{N}}}
\newcommand*{\ZZ}{{\mathbb{Z}}}
\newcommand*{\QQ}{{\mathbb{Q}}}
\newcommand*{\RR}{{\mathbb{R}}}
\newcommand*{\scrB}{\mathord{\mathcal{B}}}%
\newcommand*{\scrT}{\mathord{\mathcal{T}}}%
\newcommand*{\scrE}{\mathord{\mathcal{E}}}%
\newcommand*{\scrP}{\mathord{\mathcal{P}}}
\newcommand*{\scrS}{\mathord{\mathcal{S}}}
\newcommand*{\Shv}{\mathop{\mathrm{S\lowercase{hv}}}}

\DeclareRobustCommand\openone{\leavevmode\hbox{\small1\normalsize\kern-
.33em1}}%

\newcommand*{\EsubS}{\ensuremath{\mathord{\mathcal{E}_{\scrS}}}}
\newcommand*{\RsubC}{\ensuremath{\mathord{\RR_{\mathrm{C}}}}}
\newcommand*{\RsubD}{\ensuremath{\mathord{\RR_{\mathrm{D}}}}}

\newcommand*{\QsubD}{\ensuremath{\mathord{\QQ_{\mathrm{D}}}}}

\newtheorem{theorem}{Theorem}
\newtheorem{definition}{Definition}
\newtheorem{lemma}[definition]{Lemma}

\newtheorem{postulate}{Basic Postulate}

\begin{document}
\bibliographystyle{apsrev}

\title{\textbf{ Quantum dynamics is infinitesimal  qr-number dynamics 2}}

\author{J.V. CORBETT}
\email{john.corbett@mq.edu.au}
\address {Department of Mathematics, Macquarie University, N.S.W. 2109} %

\date{\today}

\begin{abstract} The Heisenberg equations of motion for a quantum particle of mass $m$ are deduced from the infinitesimal qr-number equations of motion for the particle. The infinitesimal qr-number equations, and hence the standard quantum mechanical equations, are related to the qr-number equations in much the same way as the equations of geometric optics are related to those of wave optics. The qr-number equations of motion for a quantum particle of mass $m$ describe the motion of a lump, given by on open set in the qr-number space of the particle, while the infinitesimal qr-number equations describe the motion of a point-like particle. The qr-number equations of motion are the Hamiltonian equations of motion for a classical particle of mass $m$ expressed in qr-numbers. The proof requires that the particle's position operators have only continuous spectrum and the force functions are smooth.

\end{abstract}



\maketitle

\section{Introduction}  The standard quantum mechanical equations of motion for a massive particle are equivalent to the infinitesimal quantum real number (aka qr-number) equations of motion. The linearity of Quantum Mechanics, as illustrated by the superposition principle, was one of the more surprising mathematical aspects of the theory. It was surprising because the equations of classical mechanics were typically non-linear. The linearisation of non-linear differential equations can be achieved using infinitesimals if the real number system contains infinitesimal numbers whose square and higher order products are zero. The standard real numbers have the Archimedean property which means that they have no infinitesimal numbers, so non-standard real numbers are required.

Quantum real numbers, introduced in \cite{5}, are not standard real numbers \cite{2}, they give numerical values to the attributes of a quantum particle dependent on its condition, see \S \ref{qrnmodel}.The qr-number equations of motion are classical Hamiltonian equations expressed in the particle's qr-numbers \cite{9}. The qr-numbers have infinitesimals whose equations of motion are shown in \S\ref{infqreq} to be Heisenberg's equations when the operators carry all the time dependence.

The qr-numbers model of quantum phenomena builds on the standard quantum mechanics, \cite{13}, using a sheaf theory extension of the standard Hilbert space formalism providing a  theory in which standard empirical results are obtained and "mysteries" of the standard interpretations are resolved. For example, the puzzle of the two slit experiment is explained in \cite{12} because the qr-number spatial continuum of a particle is topologically non-classical to the extent that a single qr-number spatial location can correspond to two separate classical locations. Furthermore, in \cite{11}, the non-point-like character of qr-number spatial locations preserves both Einstein and Bell locality. In \cite{10}, the collapse hypothesis is not needed to describe the measurement process, it gives an approximation to the qr-number value of a measured attribute.

The locally linear qr-numbers, \S \ref{llqrn}, are qr-numbers that the physical attributes of a quantum system take as their numerical values. Each is labelled by an operator $\hat A\in \mathcal{A}$, the representation of a physical attribute in the enveloping algebra of the symmetry group of the system, and like functions, each is defined on a domain that represents the conditions under which the system exists.  The conditions are complete states given by open subsets of the system's smooth state space $\mathcal{E}_{S}(\mathcal{A})$, a proper subset of the standard state space, see \S\ref{sss}. The qr-numbers are sections of the sheaf of Dedekind reals $\mathbf{R}_{\mathrm{D}}(\mathcal{E}_{S}(\mathcal{A}))$ in $\Shv(\mathcal{E}_{S}(\mathcal{A}))$, the topos of sheaves on $\mathcal{E}_{S}(\mathcal{A})$, see $\S$\ref{qrnos}.

An infinitesimal qr-number is the difference between neighbouring qr-numbers\footnote{The qr-numbers $a$ and $b$ are neighbours if there is no open subset of $\EsubS(\mathcal A )$ on which the difference between them is non-zero, or, equivalently, on which the distance between them is non-zero, see \S \ref{inflqrn}.}. For example, If $z|_{W_{0}^{\prime}}$ and $z|_{W_{0}}$ where $W_{0} \neq \emptyset$ and $W_{0}^{\prime} =W_{0} \setminus \{\rho_{0}\}$ is $W_{0}$ with a state $\rho_{0}$ removed then the difference  $z|_{W_{0}} - z|_{W_{0}^{\prime}}) = z|_{\rho_{0}} = Tr\rho_{0} \hat Z$ is an infinitesimal qr-number. There is no open subset  of $\EsubS(\mathcal{A})$ on which it is non-zero.

The infinitesimal qr-number equations of motion for a quantum particle of mass $m$ are linear equations from which the Heisenberg operator equations for the position and momentum operators of the particle can be deduced when the position operators have only continuous spectra. In brief, the standard quantum mechanical equations of motion are linear infinitesimal approximations to the qr-number equations. 
  
\section {Outline of the paper.}\label{outline} The basics of the qr-number model for a massive particle, its equations of motion and its infinitesimal qr-numbers are given in \S \ref{qrnmodel}.  Restricting attention to the equations of motion for a non-relativistic massive particle, \S \ref{equn}, we show how Heisenberg's operator equations can be obtained from the  infinitesimal qr-number equations in \S \ref{infqreq}. A proof which depends upon the position operators having only continuous spectra is given in \S \ref{ocover} and \S \ref{relb}.  

The expectation values of standard quantum mechanics are shown to be interpretable as infinitesimal qr-numbers in \S \ref{infqreq}.   The relationship between the equations of motion for infinitesimal qr-numbers and the classical equations of motion is different from that given in Ehrenfest's theorem \cite{3} which interprets the expectation values of the particle's position and momentum as standard real numbers. The proof that the equation of motion for the infinitesimal qr-number values of the particle's position and momentum are equivalent to the standard quantum mechanical equations of motion depends upon the position operators, $\{\hat Q_{j}\}_{j=1}^{3}$, having only continuous spectrum $\sigma_{c}(\hat Q_{j} =\RR$. For each $\alpha_{j} \in \sigma_{c}(\hat Q_{j})$ at least one sequence of approximate eigenstates, defined in  \S \ref{appeigen},  converges weakly to the delta function at $\alpha_{j}$, see \S \ref{examp}.

The qr-numbers, being Dedekind real numbers in a topos of sheaves on a topological space $\mathcal{E}_{S}(\mathcal{A})$, can be thought of as continuous functions defined on open subsets of $\mathcal{E}_{S}(\mathcal{A})$. The properties of the qr-numbers are given in \S \ref{qrnos}, they contain a dense subset of rational numbers given by globally constant functions on $\mathcal{E}_{S}(\mathcal{A})$, and are capable of supporting integral and differential calculus. Their underlying logic is intuitionistic, e.g. the principle of excluded middle, that $\neg \neg p \neq p$ holds,  is rejected.

The qr-number model of a non-relativistic particle of mass $m>0$, moving in an external force field, is given in \S \ref{qpart}. The infinitesimal qr-number equations of motion for its position and momentum are shown to agree with the Heisenberg equations for the position and momentum operators when the force fields which are represented by S-continuous operators,  defined in \S \ref{scfns}. This approach can be extended the two-body central force problems.

 When the S-continuous  operators $\hat F_{l} = f_{l}( \hat Q_{1},  \hat Q_{2}, \hat Q_{3}) \in \mathcal{A}$ are defined using  continuous functions $\{f_{l}\}_{l=1}^{3}$ then, at each state $\rho \in  \mathcal{E}_{S}(\mathcal{A}),$\begin{equation}\label{EQ100}
 Tr \rho \hat F_{l} = f_{l}(Tr \rho \hat Q_{1},Tr \rho \hat Q_{2},Tr \rho \hat Q_{3})\;\;\text{for}\:  l = 1,2,3.\end{equation}That is, the linear infinitesimal qr-number for $f_{j}(\vec q|_{W})$ at each state $\rho \in W $ is $Tr \rho \hat F_{j} $ for each $j =1,2,3$.
The proof in \S \ref{relb} uses approximate eigenstates for $\alpha_{j} \in \sigma_{c}(\hat Q_{j})$, the continuous spectrum of the $\{\hat Q_{j}\}_{j=1}^{3}$.

The equations of motion for the infinitesimal qr-number values of the particle's position and momentum  $ \vec q|_{\rho}$ and $\vec p|_{\rho}$ at any state $\rho \in  \mathcal{E}_{S}(\mathcal{A})$ are then given by\begin{equation}\label{EQ9}
 \frac{d}{dt}Tr \rho\hat Q_{j} = \frac{1}{m}Tr \rho \hat P_{j}\; \text{and}\; \frac{d}{dt}Tr \rho \hat P_{j} = Tr \rho \hat F_{j}.
\end{equation} Heisenberg's operator equations are deduced from these equations on the assumption that all the time dependence is borne by the operators.

Conversely starting from Heisenberg's operator equations  the qr-number equations of motion for a massive quantum particle are obtained by reversing the argument.

\section{ The qr-number model}\label{qrnmodel} The qr-number model assumes that each quantum system always has a complete state and its physical attributes retain numerical values, as qr-numbers, even when they are not being observed.  
The qr-numbers are contextual, not pure numbers, the values depends upon the physical situation of the system.

The physical attributes of a system are represented by the elements of an  O$^{\ast}$-algebra $\mathcal{A}_{S}$ on a dense subset $\mathcal{D}$ of the system's Hilbert space $\mathcal{H}_{S}$. An operator $\hat A \in \mathcal{A}_{S}$ if $\hat A : \mathcal{D} \to \mathcal{D}$ is a continuous mapping in the locally convex topology defined by the family of seminorms $\{ \| \cdot \|_{\hat A} ; \hat A \in \mathcal{A}_{S} \}$ where $\| \phi \|_{\hat A} =  \| \hat A\phi \|, \phi \in \mathcal D$, see \cite{7} for more details. The complete states are given by open subsets of the system's smooth state space and are called conditions. The smooth state space $\EsubS(\mathcal{A}_{S})$ is contained in the convex hull of projections $\scrP$ onto one-dimensional subspaces spanned by unit vectors $\phi  \in \mathcal D$.

An ambiguity in the interpretation of the standard quantum mechanics comes when, in describing an experiment, a quantum state, given by a wave function, may represent both a single system and an ensemble of systems. The qr-number interpretation avoids this inconsistency by having two classes of quantum conditions: (1) the  epistemic conditions of an ensemble of systems represent experimentally determined histories that depend on the experimenter's knowledge and (2) the ontic conditions which represents an individual system's history. Any open subset of $\EsubS(\mathcal{A}_{S})$ can be in either class but ontic conditions are always proper open subsets of an epistemic condition. The epistemic condition will always be determined by the experimental set up. The existence of ontic conditions explains the variety of outcomes. 

If an ensemble of systems has been prepared in an epistemic condition then any member of the ensemble will have an ontic condition that satisfies requirements at least as strong as those imposed by the experiment.  Mathematically, any proper open subset of an epistemic condition can be the ontic condition of a particular system in the ensemble. Ontic conditions are used to describe the evolution of individual systems, epistemic are used to describe experiments involving ensembles of "identically" prepared systems.

As in Hilbert space theories, an attribute of the system $\mathcal{S}$ is represented by an essentially self-adjoint operator $\hat A\in \mathcal{A}_{S}$, the restriction of a self adjoint operator to a dense domain $\mathcal{D}$ of the Hilbert space $\mathcal{H}_{S}$. If $\mathcal{S}$ has the condition $V \in \mathcal{O}( \EsubS(\mathcal{A}_{S}))$ then $\hat A$  has an qr-number value $a|_{V} \in \RsubD(\EsubS(\mathcal{A}_{S}))$ given by the continuous function $a: V \to \RR$ where  $\forall \rho \in V,\;a(\rho) = Tr \rho \hat A \in \Theta(\hat A)$, the numerical range of $\hat A$.\footnote{$a|_{V}$ is a representative of a section of  $\RsubD(V)$, the sheaf of germs of continuous real valued functions on $V$, where two continuous functions have the same germ at $\rho \in V$ if they agree on an open neighbourhood of $\rho$, detailed in \cite{2}.} In this paper, Greek letters are used for standard real numbers except for physical quantities like mass $m$ and a quantum state $\rho$. The qr-number  globally constant on $V$ at the standard real number $\alpha$ is written $\alpha|_{V}$.

When the O$^{\ast}$-algebra $\mathcal{A}_{S}$ is the infinitesimal representation $d\hat U$ of the enveloping algebra $\mathcal{E}(\mathcal{G})$ obtained from a unitary representation $\hat U$ of a Lie group $G$  on the Hilbert space $\mathcal{H}_{S}$. If  $d\hat U$ is the infinitesimal representation of its Lie algebra $\mathcal{G}$ on the set $\mathcal{D}^{\infty}(\hat U)$ of $C^{\infty}$-vectors of the unitary representation, then $d\hat U$ extends to a $\ast$- representation of the enveloping algebra $\mathcal{E}(\mathcal{G})$ on $\mathcal{D}^{\infty}(\hat U)$ which is dense in $\mathcal{H}_{S}$. In this situation, the O$^{\ast}$-algebra $\mathcal{A}_{S}$ is 
$d\hat U(\mathcal{E}(\mathcal{G}))$ and the open sets  $\{\nu ( \rho_{1} ; \delta)\};  \rho_{1} \in \EsubS(d\hat U(\mathcal{E}(\mathcal{G}))), \delta > 0 $ form an open basis for the weak topology on the smooth state space $\EsubS(d\hat U(\mathcal{E}(\mathcal{G})))$, corollary 8 in \cite{2}. A fuller discussion of the mathematics of the qr-number model is in \cite{2}, for our purposes it suffices to think of a qr-number in $\RsubD(\EsubS(\mathcal{A}_{S}))$ as a real-valued continuous function whose domain is an open subset of $\mathcal{A}_{S}$'s smooth state space $\EsubS(\mathcal{A}_{S})$ which is compact in the weak topology. . 

In this paper, the Hilbert space $\mathcal{H}_{S}$ of massive quantum particle carries the Schr\'odinger representation of the Weyl-Heisenberg group. $\mathcal{H}_{S} = L^{2}(\RR^{3}) = \otimes_{j=1}^{3} \mathcal H_{j}$ with each $\mathcal H_{j} = L^{2}(\RR).$ It is well known that, in this representation, all the operators $\hat Q_{j}$ and $\hat P_{j}$ are unbounded and have only continuous spectrum, see \cite{23}. The algebra $\mathcal{A}_{S}$ is the representation $d\hat U_{\hbar}(\mathcal{E}(\mathcal{G}))$ of the enveloping algebra of the Lie algebra  $h_{3}$ with basis $\{\hat P_{j}, \hat Q_{j}, \hat I \}_{j=1}^{3}$. The common domain $\mathcal{D}^{\infty}(U_{\hbar}) = \mathcal{S}(\RR^{3})$ is the Schwartz space of $C^{\infty}$-functions of rapid decrease.

 Using the trace norm topology, whose open sets are basic in the weak topology on $\EsubS(\mathcal{A}_{S}))$, both 
 \begin{equation}
 |Tr\rho_{\vec \alpha}^{n}  \hat Q_{j} - Tr \rho \hat Q_{j} | < \kappa_{m_{1}}(\hat Q_{j}) Tr |\rho_{\alpha}^{n} - \rho |\;\text{for}\; j=1,2,3 
\end{equation} and 
 \begin{equation}
 |Tr\rho_{\vec \alpha}^{n}  \hat F_{l} - Tr \rho \hat F_{l} | < \kappa_{m_{2}}(\hat F_{l}) Tr |\rho_{\alpha}^{n} - \rho | \;\text{for}\; l=1,2,3 
\end{equation} for $m_{1},\;m_{2}\in \NN^{+}$, see equation (\ref{EQ9}) in \S \ref{scfns}.

\subsection{How is a quantum particle represented?} \label{qpart}
 A quantum particle has a Hilbert space  $\mathcal{H}$ carrying an irreducible representation of its symmetry group $G$ with the Weyl-Heisenberg group as a subgroup, see \S \ref{weylheis}. Its attributes are represented by self-adjoint operators $\hat A \in \mathcal{A}$, an O$^{\ast}$-algebra of unbounded operators on $\mathcal{D}^{\infty}$ the space of $C^{\infty}$ vectors for the representation of $G$. Its smooth state space $\EsubS(\mathcal{A}) \subseteq \mathcal{E}(\mathcal{H})$ has  
pure states $\hat P_{\psi} = |\psi\rangle\langle\psi| $ for $\psi \in \mathcal{D}^{\infty} $ and mixed states are convex sums.

The condition of a particle, represented by an open subset of $\EsubS(\mathcal{A})$, gives its attributes qr-number values  with the following characteristics, generalising those of a classical particle:
\begin{enumerate}
\item the attributes are permanent but their qr-number values can vary. In experiments their qr-number values are approximated by standard rational numbers, their measured values.

\item  to a non-empty extent the particle has individuality as revealed by the qr-number values of its qualities,

\item to its non-empty extent, it can be re-identified through time using its qr-number spatial trajectory.
\end{enumerate} 

\subsection{ qr-number equations of motion for massive particles.}\label{equn} The motion of microscopic particles is governed by equations which have the same form as those for macroscopic particles with qr-numbers  replacing of standard real numbers, first postulated in \cite{12}.   

The laws of motion for a particle of mass $m>0$ are Hamiltonian equations of motion expressed in qr-numbers; \begin{equation}\label{EQ10}
m\;\frac{dq^{j}|_{U}}{ dt} = p^{j}|_{U} \;\text{and} \;
\frac{dp^{j}|_{U}}{ dt} = f^{j}(\vec q|_{U}) \end{equation} where $\hat Q^{j}$ and $\hat P^{j}$ are the $j$th components of its position and momentum and $U$ is its condition. When $h(\vec q|_{U}(t),\vec p |_{U}(t)) = \sum_{j=1}^{3} \frac{1}{ 2m}(p^{j}|_{U}(t))^{2} + V(\vec q|_{U}(t)) $ the force has components $f^{j}(\vec q|_{U}) = - \frac{\partial V(\vec q|_{U}(t))} {\partial q^{j}{U}(t)}.$ 
If the system has spin, the equations are modified. 

The time derivative of an arbitrary function $a(\vec q|_{U}, \vec p|_{U})$, when $\hat A \in \mathcal {A}$, is taken along the trajectory of the particle, is \begin{equation}\label {EQ11}
\frac{da}{dt} = [a,h] \equiv \sum ( \frac{\partial a}{\partial q_{i}} \frac{\partial h}{\partial p_{i}} -  \frac{\partial a}{\partial p_{i}} \frac{\partial h}{\partial q_{i}})
\end{equation}
The bracket $[a,h]$ is the Poisson bracket of the functions $a(\vec q|_{U}), \vec p|_{U})$ and $h(\vec q|_{U}, \vec p|_{U})$, it has the properties of a Lie bracket: skew-symmetry, $[a,b]= -[b,a]$ , bi-linearity, $[\alpha a +\gamma c,b]= \alpha[a,b]+ \gamma[c,b]$ and satisfies the Jacobi identity $[a,[b,c]] + [b,[c,a]]  +[c,[a,b]]   = 0$. 

The qr-number equation $\frac{da}{dt} = [a,h] $ is the basic dynamical equation for the evolution of the values of an attribute of a quantum system. Furthermore it shows that the conserved quantities $a(\vec q|_{U}, \vec p|_{U})$ have, with the Hamiltonian, a vanishing Poisson bracket.

\subsection{Equations for infinitesimal qr-numbers}\label{infqreq} 

The infinitesimal equations of motion are deduced from the qr-number equations using the difference between neighbouring qr-number values. 

Let $V$ be any open subset of $ \EsubS(\mathcal A)$ and put $\tilde V = V\setminus \{\rho\}$ for some state $\rho \in V$. Then \begin{equation}\label{EQ12}
\frac {d}{dt}q_{j}|_{V} - \frac{d}{dt}q_{j}|_{\tilde V} = \frac{d}{dt}Tr \rho\hat Q_{j}\; \text{and}\; \frac{d}{dt}p_{j}|_{V} - \frac{d}{dt}p_{j}|_{\tilde V} = \frac{d}{dt}Tr \rho \hat P_{j} 
\end{equation} If the component $f_{j}$ is a continuous function, \begin{equation}\label{EQ13}
 \frac{d}{dt}Tr \rho\hat Q_{j} = \frac{1}{m}Tr \rho \hat P_{j}\; \text{and}\; \frac{d}{dt}Tr \rho \hat P_{j} = f_{j}(\vec q|_{\{\rho\}})
\end{equation} because $f_{j}(\vec q|_{V})-f_{j}(\vec q|_{\tilde V}) = f_{j}(\vec q|_{\{\rho\}})$ as $\{\rho\}\cap \tilde V = \emptyset $ so that the product $q_{j}|_{\tilde V} q_{j}|_{\{\rho\}} = 0$.
For open sets $V$ and $\tilde V = V\setminus  \{\rho \}$, the difference between the qr-numbers $ f_{j}|_{V}$ and  $f_{j}|_{\tilde V}$ is  $f_{j}(Tr \rho\hat Q_{1},Tr\rho \hat Q_{2},Tr \rho \hat Q_{3})$, and that between the linear qr-numbers $\mathcal{F}_{j}|_{V}$ and  $\mathcal{F}_{j}|_{\tilde V}$ is $\mathcal{F}_{j}|_{\rho} = Tr \rho \hat F_{j}$.

Using approximate eigenvectors for the continuous spectra of the triplet of commuting position operators $(\hat Q_{1},\hat Q_{2}, \hat Q_{3})$ we show, for all states $\rho \in  \EsubS(\mathcal A)$, that \begin{equation}\label{EQ14}
Tr \rho \hat F_{l} = f_{l} (Tr \rho \hat Q_{1}, Tr\rho \hat Q_{2} ,Tr\rho \hat Q_{3} )
\end{equation} when  $\vec f :\RR^{3} \to \RR^{3}$ is $\scrS$-continuous and the $l^{th}$ component of the force operator $\hat F_{l} = f_{l}(\hat Q_{1},\hat Q_{2}, \hat Q_{3}) $  belongs to the algebra $ \mathcal{A}$.
Thus the equations for infinitesimal qr-number values of the canonical variables for a particle of mass $m$ are, for $j = 1,2,3$,\begin{equation}\label{EQ333}
 \frac{d}{dt}Tr \rho\hat Q_{j}|_{t} = \frac{1}{m}Tr \rho \hat P_{j}|_{t}\; \text{and}\; \frac{d}{dt}Tr \rho \hat P_{j}|_{t} = Tr \rho \hat F_{j}|_{t}.
\end{equation} 
Recalling that the standard unitary quantum dynamics is given \begin{itemize} \item by Schr\"odinger's equation for wave-functions, vectors $\psi \in \mathcal{H}$, or 
\item by Heisenberg's equations for operators $\hat A$ representing physical quantities. \end{itemize}The unitary operator is $\hat U_{t} = \exp -\frac{\imath t}{\hbar}\hat H$ for the self-adjoint Hamiltonian operator, $\hat H$, the unitary time evolution is $\hat A_{t} = \hat U_{t} ^{-1} \hat A \hat U_{t} $ for any operator $\hat A$.  Heisenberg's equations for a Galilean relativistic particle of mass $m$, when the Hamiltonian $\hat H = \frac{1}{2m}\sum _{j=1}^{3}\hat P_{j}^{2} + V(\hat Q_{1},\hat Q_{2},\hat Q_{3})$  are, for $ j = 1,2,3$, \begin{equation}
\frac{d \hat Q_{j}|_{t} }{dt} = \frac{1}{m}\hat P_{j}|_{t}  \;\; \text{and} \;\; \frac{d \hat P_{j}|_{t}} {dt} =  f_{j}(\hat Q_{1}|_{t} ,\hat Q_{2}|_{t} ,\hat Q_{3}|_{t} ) = \hat F_{j}|_{t}
\end{equation} where  $f_{j} = -\frac{\partial V}{\partial q_{j}}$ is the force function. 

If the operators are assumed to carry all the time dependence in the infinitesimal qr-number equations (\ref{EQ333}), then \begin{equation}
 Tr \rho  \frac{d}{dt}\hat Q_{j}|_{t} = \frac{1}{m}Tr \rho \hat P_{j}|_{t}\; \text{and}\; Tr \rho  \frac{d}{dt}\hat P_{j}|_{t} = Tr \rho \hat F_{l}|_{t}\;\; \text{for}\;  j = 1,2,3
\end{equation} which holds for all $\rho \in \EsubS(\mathcal A)$ so that the operators satisfy Heisenberg's equations.

\subsection{The evolution of the conditions}\label{evolcon} Conversely, if the time dependence in the infinitesimal qr-number equations (\ref{EQ333}) is assumed to be borne by the states then \begin{equation}
\rho_{t} = \hat U_{t} \rho \hat U_{t}^{-1}
\end{equation}

Any condition will evolve through the unitary evolution of each of its component states that is if $\rho_{0} \to \rho_{t} = \hat U_{t} \rho_{0} \hat U_{t}^{-1}$ for all $\rho_{0} \in W_{0}$ then $ W_{0} \to W_{t} =   \hat U_{t} W_{0} \hat U_{t}^{-1}$.  This demonstrated by first showing that $\nu(\rho_{0}, \delta) \to  \nu(\rho_{t}, \delta)$, then by using the fact that  the open sets $\{\nu(\rho, \delta)\}$ are basic in the topology on state space, that is, that every open set is a union of sets of the form $\nu(\rho, \delta)$.
 \begin{lemma}\label{L7} If $\rho, \rho^{\prime}\in \EsubS(\mathcal{S})$ then $Tr|\rho_{t} - \rho^{\prime}_{t}| = Tr|\rho_{0} - \rho^{\prime}_{0}|$ when $\rho_{t} = \hat U_{t} \rho \hat U_{t}^{-1}$ for a unitary group $\{ \hat U_{t}; t \in \RR \}$, thus if $\rho_{0}\to \rho_{t}$ then $\nu(\rho_{0}, \delta) \to  \nu(\rho_{t}, \delta)$ for any $\rho_{0}$, any $\delta >0$ and unitary group $\hat U_{t}$.
\end{lemma} The proof uses the result, $| \hat U_{t} (\rho - \rho^{\prime} ) \hat U_{t}^{\ast}| = \hat U_{t}|\rho - \rho^{\prime} | \hat U_{t}^{\ast}$, and that the trace is independent of the orthonormal basis used in its evaluation.

\subsection{Afterwords}\label{after}
 The attributes of individual microscopic systems in an experiment always have qr-number values so that individual systems can be followed. A quantum particle with non-zero mass moves along a trajectory in its qr-number space;  its position and momentum values $ (\vec q|_{U}(t),\vec p|_{U}(t)) $ satisfy Hamilton's equations of motion. The location of the particle in in its spatial continuum is an open subset, a lump, not a point. Therefore a location need not be connected and can be sub-divided. This feature, together with the absence of the law of excluded middle in its logic, allows the qr-number model to have a different understanding of phenomenon like the double slit experiment, \cite{15}, and Einstein and Bell non-locality,\cite{14}. The infinitesimal equations of motion provide a more point-like description of the trajectory.
 
Although spin has not been explicitly discussed above, using the irreducible projective unitary representations of the Galilean group $G$, see \cite{5}, the quantum mechanical equations for particles with spin are obtained as infinitesimal approximations. 

These representations are labeled by the mass $m$, internal energy $U$ and spin $s$, where $m > 0$  and
$U$ are standard real numbers and $s$ is an integer or a half integer. The representation labeled $(m, U, s)$ acts on the Hilbert
space $\scrH :=
\mathcal{L}^{2}(\mathbb{R}^3) \otimes \mathbb{C}^{2s +1}$. The elements of
$\scrH$ are $(2s + 1)$-component vectors of square integrable functions
$\vec x \in \RR^{3} \mapsto \{ \psi_{i} (\vec x) \}, i = -s,..,s $. In the representation $(m, U, s)$ with $m>0$, the position operators $\hat Q_j,  j =1,2,3,$ have only continuous spectrum with $\sigma_{c}(\hat Q_{j}) =\RR$, hence the argument that Heisenberg equations give an infinitesimal approximation to the qr-number equations still holds. 
 
 The qr-number equations of motion for a quantum particle of mass $m$ are the classical Hamiltonian equations of motion for a particle of mass $m$ expressed in qr-numbers.  These equations, being applicable to both classical and quantum Galilean relativistic mechanics, exemplify Einstein's statement, "I do not believe in micro and macro laws, but only in (structural) laws of general validity", \cite{1} pp 141. While the algebraic structure of the equations remains the same for both the macro and micro Galilean relativistic worlds, different systems of real numbers are taken as the values of the entities in the different worlds. \begin{postulate} \label{laws}The general laws of physics should be expressed by equations which hold good for all systems of real numbers.
\end{postulate} The qr-number model of quantum physics provides an example that supports the hypothesis that the general laws of physics should be expressed by equations which hold good for all systems of real numbers. 
Different physical systems may take different systems of real numbers as the numerical values of their physical attributes.

If the universe of real number systems is assumed to be that of Dedekind real numbers constructed in a topos of sheaves on a topological space $E$, see \cite{10} \S VI.8, the different systems of real numbers would come from different families of continuous functions defined on the base space $E$. A system's real numbers would then be continuous functions defined on non-empty open subsets of its state space $E$. The properties of such systems of real numbers would mimic those of qr-numbers in \S \ref{qrnos}, they would all contain rational numbers as a dense subset and be capable of supporting integral and differential calculus. Their underlying logic would be intuitionistic and Boolean when the weak topology on $E$ is the weakest topology on $E$, namely when the only open sets are $( \emptyset, E )$. This will occur when all the functions are globally constant maps from $E \to \RR$ which are continuous is the indiscrete topology.  For if $\gamma \in \RR$ is the map constant at $\gamma$ then the inverse image $\gamma^{-1}(J)$ of any open subset $J \subset \RR$ is $\emptyset$ if $\gamma \notin J$ and is $E$ if $\gamma \in J$.

This discussion must be modified for special relativistic systems. Even when explicit formulae for the operators $\hat X_{\mu}$  that represent the the points in the Minkowski space-time are obtained more is needed., For example in Jaekel and Reynaud work \cite{4} that uses representations of the conformal group, the  operators  $\hat X_{\mu}$ all don't commute for non-zero spin.

\section{Appendices}
\subsection{O$^{\ast}$-algebras}
If $\mathcal D$ is a dense subset of a
Hilbert space $\scrH$, let $\mathcal L(\mathcal D)$ denote the 
set of all linear operators from $\mathcal D$ to $\mathcal D$. 
If $\hat A^{\ast}$ is the Hilbert space adjoint of a linear operator $\hat A$ whose domain, $\text{dom}(\hat A) \supseteq\mathcal{D}$, put
\begin{equation} 
\mathcal{L}^{\dagger}(\mathcal D) = \{ \hat A \in \mathcal L(\mathcal D) ; \text{dom}(\hat A^{\ast})
\supset \mathcal D,  \hat A^{\ast} \mathcal D \subset \mathcal D\}
\end{equation}
Then $\mathcal L^{\dagger}(\mathcal D) \subset  \mathcal L(\mathcal D) $ where $ \mathcal L(\mathcal D) $ is an algebra with the usual operations for linear operators with a common invariant domain: addition $\hat A +\hat  B$, scalar multiplication $\alpha \hat A$ and
non-commutative multiplication $\hat A\hat B$. Furthermore $\mathcal L^{\dagger}(\mathcal D)$  is a $\ast$-algebra with the involution $\hat A \to \hat A^{\dagger} :\equiv \hat A^{\ast} |_{\mathcal D}$. \begin{definition}
An  O$^{\ast}$-algebra on $\mathcal D \subset \scrH$ is a $\ast$-subalgebra of $\mathcal L^{\dagger}(\mathcal D)$. 
\end{definition}  The locally convex topology defined by the family of seminorms $\{ \| \cdot \|_{\hat A} ; \hat A \in \mathcal A \}$ where $\| \phi \|_{\hat A} =  \| \hat A\phi \|, \phi \in \mathcal D$. It is the weakest locally convex topology on $\mathcal D$ relative to which each operator $\hat A \in \mathcal A$ is a continuous mapping of $\mathcal D$ into itself, see Schm\"udgen,\cite{7}.

When $\hat U$ is a unitary representation of a Lie group $G$ on a Hilbert space $\mathcal{H}$ with $d\hat U$ the infinitesimal representation of its Lie algebra $\mathcal{G}$ on the set $\mathcal{D}^{\infty}(\hat U)$ of $C^{\infty}$-vectors, the O$^{\ast}$-algebra $\mathcal{A}$, used in the construct of the qr-numbers,  is the infinitesimal representation $d\hat U$ of the enveloping algebra $\mathcal{E}(\mathcal{G})$. For the Schr\"odinger representation of the enveloping Lie algebra of the Weyl-Heisenberg group, see.\S \ref{weylheis}, $\mathcal{D}^{\infty}(U) = \mathcal{S}(\RR^{3})$,  the Schwartz space of $C^{\infty}$-functions of rapid decrease. 

\subsection{States on an O$^{\ast}$-algebra $\mathcal A$}\label{sss}  Denote by $\scrT_{1}(\scrH)$ the collection  of all trace class operators on $\scrH$. For the $C^{\ast}$-algebra $ \scrB( \scrH)$, the state space $\scrE = \EsubS(\scrB( \scrH))$ is composed of operators in $\scrT_{1}(\scrH)$ that are self-adjoint and normalised. 

 Define a subset $\mathcal{T}_{1}(\mathcal A) \subset \scrT_{1}(\scrH)$ composed of trace class operators $\hat T$ that satisfy $ \hat T\scrH \subset \mathcal D , \hat T^{\ast}\scrH \subset \mathcal D $ and $\hat A \hat T, \hat A \hat T^{\ast} \in \mathcal{T}_{1}(\scrH)$ for all $\hat A \in \mathcal A $, that is,
$\mathcal{T}_{1}(\mathcal A)$ is a $\ast$-subalgebra of $\mathcal A$ satisfying $\mathcal A
\mathcal{T}_{1}(\mathcal A) = \mathcal{T}_{1}(\mathcal A)$\cite{7}. 
\begin{definition}
The smooth state space $\EsubS(\mathcal A)$ for the $O^{\ast}$-algebra $\mathcal A$ is the set of normalized, self-adjoint operators in $\mathcal{T}_{1}(\mathcal A)$.
\end{definition}
\subsection{ Mathematics of qr-numbers}\label{qrnos} qr-numbers are sections of the sheaf of Dedekind reals $\RsubD(\EsubS(\mathcal{A}))$ in $\Shv(\EsubS(\mathcal{A}))$,defined in \cite{7} and \cite{8}. $\Shv(\EsubS(\mathcal{A}))$ is a topos of sheaves on the smooth state space $\EsubS(\mathcal{A})$. 

$\EsubS(\mathcal{A})$ has the weak topology, generated by the functions $ a(\cdot)$ defined for each $\hat A \in  \mathcal A,  \;a( \rho) = Tr \hat A  \rho,  \forall   \rho  \in \EsubS(\mathcal A) $. This is the weakest topology making all the functions $ a(\cdot)$ continuous, in it $\EsubS(\mathcal A)$ is compact. With parameters $\hat A \in \mathcal A, \epsilon > 0$ and $\rho_0 \in \EsubS(\mathcal A)$, the sets  $ \scrN( \rho_{0}  ; \hat A ; \epsilon) = \{
\rho \ ; |Tr \rho \hat A - Tr \rho_{0} \hat A| < \epsilon \}$  form an open 
sub-base for the weak topology on $\EsubS(\mathcal A)$. When $\mathcal{A}$ is the infinitesimal representation $d\hat U$ of the enveloping algebra $\mathcal{E}(\mathcal{G})$ on the set $\mathcal{D}^{\infty}(\hat U)\subset \mathcal {H}$ of $C^{\infty}$-vectors obtained from a unitary representation $\hat U$ of a Lie group $G$ on $\mathcal{H}$ then the open sets  $\{\nu ( \rho_{1} ; \delta) = Tr |\rho - \rho_{1}|< \delta \};  \rho_{1} \in \EsubS(\mathcal{A}), \delta > 0 $ form an open basis for the weak topology on $\EsubS(\mathcal{A})$. For more details see \cite{2}.

$\RsubD(\EsubS(\mathcal{A}))\cong \mathcal C(\EsubS(\mathcal{A}))$, the sheaf of germs of continuous real-valued functions on $\EsubS(\mathcal{A})$. Two functions have the same germ at $\rho \in \EsubS(\mathcal{A})$ if they agree on an open set containing $\rho$.   

On any non-empty set $U \in \mathcal{O}( \EsubS(\mathcal{A}))$, the subsheaf $\RsubD(U)$ of qr-numbers:
\begin{itemize}
\item has integers $\ZZ(U)$, rationals 
$\QQ(U)$ and Cauchy reals $\RsubC(U)$ as subsheaves of 
locally constant functions; 
\item has orders $<$ and  $\leq$ compatible with those on $\QQ(U)$  but the inequality $<$ is not total because trichotomy, $x>0 \;\lor \;x=0\; \lor \;x<0$, is not satisfied and $\leq\;\;\not\equiv\;\;<\; \lor\;=$; 
\item is closed under the commutative, associative,
distributive binary operations  $+$  and $\times$, has $0 \ne 1$
and is a residue field, i.e., if $ b \in \RsubD(U)$ is not invertible then $b = 0$; 
\item has a distance function, $| a|_{U}| = max(a|_{U}, -a|_{U})$, which defines a pseudo-metric with respect to which $\RsubD(U)$ is a complete pseudo-metric space in which $\QQ(U)$ is dense. 
A section $b\in \RsubD(U)$ is apart from $0$ iff  $|b| > 0$. $\RsubD(U)$ is an apartness field, i.e., $\forall b \in \RsubD(U), \;  |b| > 0 $ iff $b$ is invertible.
\item is not Dedekind complete: bounded sets need not have least upper bounds and is not Archimedean as it has infinitesimals, e.g. each expectation value, such as $a|_{\rho} = Tr \rho \hat A$,  is an  intuitionistic nilsquare infinitesimal, as $\lnot [a|_{\rho}^2 \ne 0]$ is true and an order theoretical infinitesimal qr-number because there is no open set on which $ a|_{\rho} > 0 \lor a|_{\rho} < 0 $ is true.. 
\end{itemize}
\subsection{Locally linear qr-numbers}\label{llqrn}

The locally linear qr-numbers are a subsheaf $\mathbf{A}(\mathcal{E}_{S}(\mathcal{A}))$ of $\mathbf{R}_{\mathrm{D}}(\mathcal{E}_{S}(\mathcal{A}))$. Given an essentially self-adjoint or symmetric $\hat A \in \mathcal{A}$ there is a locally linear real-valued function $a$ on  $\mathcal{E}_{S}(\mathcal{A})$ with $a(\rho) = Tr \rho \hat A$ for $\rho \in \mathcal{E}_{S}(\mathcal{A})$. Their properties include:
\begin{itemize} 
 \item $\mathbf{A}(\mathcal{E}_{S}(\mathcal{A})))$ is dense in $\mathbf{R}_{\mathrm{D}}(\mathcal{E}_{S}(\mathcal{A}))$.  
 \item Every qr-number is a continuous real function of locally linear qr-numbers.
\item If $W \neq \emptyset$, $a|_{W} = b|_{W}$ if and only if $\hat A = \hat B$.
\end{itemize}
The sheaf of locally linear qr-numbers
$\AAA(\EsubS(\mathcal A))$  is dense in $\RsubD(\EsubS(\mathcal A))$ in the metric
topology $T$ because they include the locally constant functions with rational values which form a sub-sheaf $\QsubD(\EsubS(\mathcal A))$ of $\AAA(\EsubS(\mathcal A))$ which is dense in $\RsubD(\EsubS(\mathcal A))$ in the metric topology $T$\cite{8}.
\begin{lemma}\label{LEM44}
Given a qr-number $b|_{U}$ on the open set $U\in \EsubS(\mathcal A)$ and $\epsilon >0$ there exists an open cover $\{U_{j}\}$ of $U$ such that for each $j$ there is a locally linear function $a : U_{j} \to \RR$ such that $|b|_{U_{j}} - a|_{U_{j}}| < \kappa\epsilon$, for a finite constant $\kappa$ of the same physical dimensions as $b$ and $a$.
\end{lemma}

\subsection{Infinitesimal  qr-numbers}\label{inflqrn} 

Infinitesimal  qr-numbers are non-zero functions $\EsubS(\mathcal A) \to \RR$  for which there is no non-empty open subset of $\EsubS(\mathcal A)$  on which they are non-zero. That is, they are only non-zero on sets with an empty interior. The infinitesimal qr-numbers are constructed as a sheaf on $\EsubS(\mathcal A)$ by prolonging by zero a section defined on a subset of $\EsubS(\mathcal A)$ with empty interior, c.f. \cite {17}.

The section of a continuous function $f$ defined on a non-empty set $S \subset \EsubS(\mathcal A)$ is an order theoretical infinitesimal number if there is no non-empty open set $V\subseteq \EsubS(\mathcal A)$ on which $f|_{V} > 0 \lor f|_{V} < 0$. They originally arise as the difference between neighbouring qr-numbers. 

The qr-numbers $a$ and $b$ are neighbours if there is no open subset of $\EsubS(\mathcal A )$ on which the difference between them is non-zero, or, equivalently, on which the distance between them is non-zero. Therefore $a$ and $b$ are neighbouring if $a \neq b$ and there is no open set on which the difference between them is non-zero,\begin{equation}
  \text{ neither} \; (a - b)|_{V} < 0 \;\text{nor}\; (a - b)|_{V} > 0 \text{ for any}\; V \neq \emptyset.
\end{equation} 

 For example, if $V_{0} = \nu (\rho_{0};\delta )$, a basic open set in the weak topology on  $\EsubS(\mathcal{A })$, for  $\rho_{0}\in \EsubS(\mathcal A )$ and $\delta >0$, and $\tilde V_{0} = V_{0}\setminus \{\rho_{0}\}$,  then the qr-numbers $a|_{V_{0}} $ and $a|_{\tilde V_{0}}$ are neighbours because  \begin{equation}
a|_{V_{0}} \neq a|_{\tilde V_{0}},\; \text{but neither} \; a|_{V_{0}} < a|_{\tilde V_{0}}\;\text{nor}\; a|_{\tilde V_{0}} < a|_{V_{0}}
\end{equation} on any open subset of $ \EsubS(\mathcal A )$ for  $a|_{V_{0}} - a|_{\tilde V_{0}} = a|_{\rho_{0}} $ and the singleton set $\{\rho_{0}\}$ has empty interior. 

When the qr-numbers are locally linear, $ a|_{\rho_{0}} = Tr \rho \hat A$ for some essentially self-adjoint $\hat A \in \mathcal{A}$ and $\rho \in V_{0}$ and $\rho_{0} = \hat P_{\phi_{0}}$ is pure then the expectation value $Tr\rho_{0} \hat A = (\phi, \hat A \phi)$  of standard quantum mechanics is interpretable as an order theoretic infinitesimal qr-number.

Given the  qr-number $f|_{U}$ Lemma \ref{LEM44} in \S \ref{llqrn} asserts that there is an open cover $\{U_{j}\}$ of $U$ on each component of which there is a locally linear function $a|_{U_{j}}$ that approximates $f|_{U_{j}}$ arbitrarily well.
Since each $\rho \in U$ belongs to an open set $U_{k}\in \{U_{j}\}$, for each qr-number $f|_{U}$, $\exists \hat A \in \mathcal{A}$ such that the infinitesimal qr-number $f|_{\rho}  = Tr \rho \hat A$.

\section{Proofs} 

\subsection{Basic results }\label{scfns}

A function $\vec f :\RR^{3} \to \RR^{3}$ is continuous 
if each component $f_{j} , \; j =1,2,3$,  is a continuous function, $\RR^{3} \to \RR$. \begin{definition}\label{D3}
 If as a function of position operators  $\{\hat Q_{j}\}_{j=1}^{3},$ each function $f_{j}$ defines an essentially self-adjoint operator, $\hat F_{j} = f_{j}(\hat Q_{1},\hat Q_{2},\hat Q_{3} )$ mapping $\mathcal{S}(\RR^{3}) \to \mathcal{S}(\RR^{3})$, it is called an $\scrS$-continuous operator.
\end{definition}
The $\scrS$-continuous operator $\hat F_{j} = \sum_{|n| = 1}^{N}a(j)|_n \hat Q_{1}^{n_{1}}\hat Q_{2}^{n_{2}}\hat Q_{3}^{n_{3}},$ with multi-index $n = (n_{1},n_{2},n_{3})$, $|n| =  n_{1}+n_{2} +n_{3}$ for non-negative integers $n_{j}$ and $N <\infty,$ may be associated with two qr-numbers: a non-linear qr-number,  $ f_{j}|_{W} = \sum_{|n| = 1}^{N}a(j)|_n  (q_{1}|_{W})^{n_{1}}(q_{2}|_{W})^{n_{2}}(q_{3}|_{W})^{n_{3}}$, or a linear qr-number $\mathcal{F}_{j}|_{W}$ for the operator $\hat F_{j}$ with $\mathcal{F}_{j}(\rho) = Tr \rho \hat F_{j}, \; \forall \rho \in W$. 

For the non-linear qr-number,  each $f_{l} : \RsubD(\EsubS(\mathcal{A}_{S}))^{3} \to \RsubD(\EsubS(\mathcal{A}_{S})) $ takes a triplet of locally linear qr-numbers $\vec q|_{W}$ to a qr-number $f_{l}(\vec q|_{W})$. The $f_{l}$  are continuous in the pseudo-metric topology of the non-negative function $d :\RsubD(\EsubS(\mathcal{A}_{S})) \times \RsubD(\EsubS(\mathcal{A}_{S})) \to \RsubD(\EsubS(\mathcal{A}_{S}))^{+}$ with $d(a,b) = \max (a-b,b-a)$ for any pair of qr-numbers $a,b$, see \S \ref{qrnos}. 

Also, for any essentially self-adjoint operator $\hat A \in \mathcal{A}_{S}$, the function mapping $\rho \in \EsubS(\mathcal{A}_{S}) \to Tr \rho \hat A \in \RR$ is continuous in the trace-norm topology on $\EsubS(\mathcal{A}_{S})$ and the standard metric topology on $\RR$.  Using the trace norm topology we have for any $\rho, \rho_{0} \in \EsubS(\mathcal{A}_{S})$,
 \begin{equation}\label{EQ9}
 |Tr\rho_{0}  \hat A - Tr \rho \hat A | < \kappa_{m}(\hat A) Tr |\rho_{0}- \rho |
\end{equation} for $m \in \NN^{+}$, because for any such $\hat A \in \mathcal{A}, \exists m \in \NN^{+}$  so that \begin{equation}\label{EQ110}
\kappa_{m}(\hat A) = \sup_{\psi \in \mathcal{D}^{\infty}(\hat U)} \| \hat A d\hat U((1- \Delta)^{m})^{-1} \psi \| / \|\psi\| <\infty \end{equation} where $\Delta = \sum_{i=1}^{d} x_{i}^{2}$ is the Nelson Laplacian in the enveloping algebra of the Lie algebra $\mathcal{G}$ with basis $\{ x_{1},x_{2},......,x_{d}\}$ and $d\hat U(\Delta)$ is its representative derived from the unitary representation $\hat U$ of the Lie group $G$, proved in \S E of \cite{2}. In this paper, $G$ is the Heisenberg-Weyl group and in the Schr\"odinger representation the basis elements of the enveloping algebras are $\{\hat P_{j}, \hat Q_{j}, \hat I \}_{j=1}^{3}$ so that $d\hat U(\Delta) = \sum_{j=1}^{3} (\hat P_{j}^{2} + \hat Q_{j}^{2} )+\hat I $.

In fact, because $\EsubS(\mathcal{A}_{S})$ is compact in the weak, and hence the trace-norm, topology these functions are uniformly continuous which means that for every $\epsilon >0$ there exists a $\delta >0$ such that \begin{equation}
 |Tr\rho_{0}  \hat A - Tr \rho \hat A | < \epsilon \; \forall\; \rho, \rho_{0} \in \EsubS(\mathcal{A}_{S}) \;\text{with}\; Tr |\rho_{0}- \rho |< \delta.
\end{equation}
The function that sends \begin{equation}
\rho \in \EsubS(\mathcal{A}_{S}) \to f_{l}(Tr \rho \hat Q_{1},Tr \rho \hat Q_{2},Tr \rho \hat Q_{3}) \in \RR
\end{equation} is also uniformly continuous when $\EsubS(\mathcal{A}_{S})$ has the trace-norm topology and $\RR$  its standard metric topology.

\subsection{Approximate eigenstates for continuous spectrum}\label{appeigen}
Consider the operator $\hat Q_{j}$, the $j$th component of the position operator. If $\lambda_{j} \in \RR$ is in $\sigma_{c}(\hat Q_{j})$, the continuous spectrum of $\hat Q_{j}$, then there is a sequence of vectors $\{\phi^{n}_{\lambda_{j}}\}_{n=1}^{\infty}$ in the domain of $\hat Q_{j}$ such that $\lim_{n \to \infty}\| \hat Q_{j} \phi^{n}_{\lambda_{j}} - \lambda_{j} \phi^{n}_{\lambda_{j}}\| =0$, even though $\{\phi^{n}_{\lambda_{j}}\}_{n=1}^{\infty}$  contains no strongly convergent subsequence. This is the Weyl criterium for continuous spectrum \cite{15}, the vectors are called approximate eigenvectors.
The corresponding sequence of approximate eigenstates is given by the projection operators $\{ \hat P_{\phi^{n}_{\lambda_{j}}}\}_{n=1}^{\infty}$, that project onto the rays of vectors $\phi^{n}_{\lambda_{j}}\in \mathcal{D}(\hat Q_{j})$.

If $\rho_{\vec \alpha}^{n} = \hat P_{\Phi_{\vec \alpha}^{n}}$ for $\vec \alpha =(\alpha_{1}, \alpha_{2},\alpha_{3})$ with $\alpha_{j}  \in \sigma_{c}(\hat Q_{j})$, the continuous spectrum of $\hat Q_{j}$, and  $\Phi_{\vec\alpha}^{n} = \otimes_{j=1}^{3} \phi_{\alpha_{j}}^{n}$ where $\{\phi_{\alpha_{j}}^{n}\}_{n=1}^{\infty} \in \mathcal{D}(\hat Q_{j})$ is a sequence of approximate eigenvectors for $\hat Q_{j}$ at $\alpha_{j}$ then $\forall \delta>  0$, $\exists N_{0}( \delta)$\footnote{$N_{0}(\delta)$ is independent of $\vec \alpha \in \RR^{3}$, see \S\ref{examp}. } such that, for each $j =1,2,3,$ if  $n > N_{0}( \delta)$, 
 \begin{equation}\label{EQ43}
\| (\hat Q_{j} - \alpha_{j})\phi_{\alpha_{j}}^{n}\| < \delta \;\; \text{and}\;\;  |Tr \rho_{\vec \alpha}^{n}\; \hat Q_{j} - \alpha_{j}| < \delta.\end{equation} 

From the Lemmas (\ref{L11}) and (\ref{L12}) below, which are demonstrated using the  Schr\"odinger representation of the  enveloping algebra of the Weyl-Heisenberg Lie algebra in \S \ref{weylheis}, it follows that  \begin{equation}
\lim _{n\to \infty}|Tr \rho^{n}_{\vec \alpha} f_{j}(\hat Q_{1},\hat Q_{2},\hat Q_{3}) - f_{j}(Tr \rho^{n}_{\vec \alpha} \hat Q_{1},Tr \rho^{n}_{\vec \alpha} \hat Q_{2},Tr \rho^{n}_{\vec \alpha} \hat Q_{3})| = 0 \end{equation}
 as both $Tr \rho^{n}_{\vec \alpha} f_{j}(\hat Q_{1},\hat Q_{2},\hat Q_{3})$ and $f_{j}(Tr \rho^{n}_{\vec \alpha} \hat Q_{1},Tr \rho^{n}_{\vec \alpha} \hat Q_{2},Tr \rho^{n}_{\vec \alpha} \hat Q_{3})$ tend to $f_{j}(\vec \alpha)$ as $n \to \infty$.

\begin{lemma}\label{L11}
For any triplet of real numbers  $\vec \alpha$ and triplet of $\scrS$-operators $\hat F_{j} =  f_{j}(\hat Q_{1},\hat Q_{2},\hat Q_{3} )$, if $\{\rho^{k}_{\vec \alpha} = \hat P_{\Phi^{k}_{\vec \alpha}}  \}$ are the  joint approximate eigenstates 
 for the operators $\{\hat Q_{j}\}_{j=1}^{3}$  then
\begin{equation}
\lim_{k \to \infty} Tr\rho^{k}_{\vec \alpha}\hat F_{j}=f_{j}( \vec \alpha).
\end{equation} 
\end{lemma}
The $\rho^{k}_{\vec \alpha} = \hat P_{\Phi^{k}_{\vec \alpha}} $ are pure states, projection operators onto the one-dimensional subspaces spanned by the unit vectors $\Phi^{k}_{\vec \alpha}$. For any continuous function $f_{j}(\vec x)$ from $\RR^{3}$ to $\RR$, in the limit as $k \to \infty$, 
\begin{equation}
Tr \rho^{k}_{\vec \alpha} f_{j}(\hat Q_{1},\hat Q_{2},\hat Q_{3}) =  \int f_{j}(\vec x)|\Phi^{k}_{\vec \alpha}( \vec x - \vec \alpha)|^{2} d^{3}x \to f_{j}(\vec \alpha)
\end{equation} because  $|\Phi^{k}_{\vec \alpha}( \vec x - \vec \alpha)|^{2}  \to \delta (\vec x - \vec \alpha)$, c.f. examples in \S \ref{examp}. That is, for any $\epsilon >0$, there is an integer $N_{1}(\epsilon)$ such that for all $k > N_{1} (\epsilon)$,
\begin{equation}\label{EQ44}
|Tr \rho^{k}_{\vec \alpha}\hat F_{j} - f_{j}( \vec \alpha)| <  \frac{\epsilon}{6}.
\end{equation}

\begin{lemma}\label{L12}   For each $l$, $|f_{l}(Tr \rho^{k}_{\vec \alpha} \hat Q_{1},Tr \rho^{k}_{\vec \alpha} \hat Q_{2},Tr \rho^{k}_{\vec \alpha} \hat Q_{3}) -Tr \rho^{k}_{\vec \alpha}\hat F_{l}|$  approaches zero as $k \to \infty$,  when  the $\rho^{k}_{\vec \alpha} $ are joint approximate eigenstates for the operators $\{\hat Q_{j}\}_{j=1}^{3}$ at $\{\alpha_{j}\}_{j=1}^{3}$.\end{lemma} 

For $l = 1,2,3$, $Tr \rho^{k}_{\alpha_{l}} \hat Q_{l} \to \alpha_{l}$ as $k \to \infty$ and each component $f_{j}$ is a continuous function from $\RR^{3}$ to $\RR$, then for any $\vec \alpha \in \RR^{3}$ and $\epsilon >0$, there is $\delta_{j} >0$ and $N_{0}( \delta_{j})$ such that $n >N_{0}( \delta_{j})$ implies   $ |Tr \rho_{\vec \alpha}^{n}\; \hat Q_{j} - \alpha_{j}| < \delta_{j}$ then for all $k > N_{0}( \delta_{j}) $,
\begin{equation}\label{EQ441}
| f_{j}(Tr \rho^{k}_{\vec \alpha} \hat Q_{1},Tr \rho^{k}_{\vec \alpha} \hat Q_{2},Tr \rho^{k}_{\vec \alpha} \hat Q_{3}) - f_{j}(\vec \alpha)| < \frac{\epsilon}{6}.
\end{equation}
Thus, given $\epsilon > 0$, for each $\vec \alpha \in \RR^{3}$ and all  $k >\max(N_{0}(\delta_{j}),N_{1}(\epsilon))$, \begin{equation}\label{EQ45}
|Tr \rho^{k}_{\vec \alpha}\hat F_{j} - f_{j}(Tr \rho^{k}_{\vec \alpha} \hat Q_{1},Tr \rho^{k}_{\vec \alpha} \hat Q_{2},Tr \rho^{k}_{\vec \alpha}\hat Q_{3})| < \frac{\epsilon}{3}.
\end{equation}  
Hence for $k >\max(N_{0}(\delta_{j}),N_{1}(\epsilon))$, $f_{j}(Tr \rho^{k}_{\vec \alpha}\hat Q_{1},Tr\rho^{k}_{\vec \alpha} \hat Q_{2},Tr \rho^{k}_{\vec \alpha} \hat Q_{3}) = Tr \rho^{k}_{\vec \alpha}\hat F_{j}$ as infinitesimal qr-numbers when each $\hat Q_{j},$$j =1,2,3$ has
only continuous spectrum.

\subsection{Representation of the Weyl-Heisenberg group}\label{weylheis}

The Heisenberg Lie algebra $h_{3}$ for a $3$ dimensional quantum system has a basis $\{ X_{j}, Y_{j}, Z\}_{j=1}^{3}$ of skew-hermitian operators.  The Lie brackets, using the Kronecker delta $\delta_{j,k}$, are \begin{equation}
[X_{j},Y_{k}] = Z \delta_{j,k},\;\;\; [X_{j},Z] = 0,\;\;\; [Y_{j},Z] = 0.\end{equation}

In Schr\"odinger's representation $\hat U_{\hbar}$ of the Weyl-Heisenberg Lie algebra on $ \mathcal{D}^{\infty}(U_{\hbar}) = \mathcal{S}(\RR^{3}) \subset L^{2}(\RR^{3})$,  $ \mathcal{S}(\RR^{3})$ is  the Schwartz space of $C^{\infty}$-functions of rapid decrease. 
 It is well known that if we take, for $j =1,2,3$,  $ X_{j} = \imath \hat P_{j}$,  $ Y_{j} = \imath \hat Q_{j}$ and  $ Z = \imath\hbar \hat I$, then in the Schr\"odinger representation, for  $j =1,2,3$ and all $\phi(\vec x)\in \mathcal{S}(\RR^{3})$,\begin{equation}
 \hat Q_{j} \phi(\vec x) =  x_{j}\phi(\vec x)\; \text{and}\; \hat P_{j} \phi(\vec x) = - \imath  \hbar \frac{d}{dx_{j}}\phi(\vec x).\end{equation}  Each operator $\hat Q_{j}$ and $\hat P_{j}$ is unbounded and has only continuous spectrum, see \cite{16}, and they satisfy the canonical commutation relations, \begin{equation}\label{CCR}
[\hat Q_{j} , \hat P_{k}] = \imath \hbar \delta_{jk}\;\;\text{and}\;\; [\hat Q_{j} , \hat Q_{k}] = 0, \;\;  [\hat P_{j} , \hat P_{k}] = 0
\end{equation} for $j$ and $k$ equal to $1,2,3$.

The O$^{\ast}$-algebra $\mathcal{A}$, that we use to construct the qr-numbers for Schr\"odinger's representation of the Weyl-Heisenberg Lie algebra,  is the representation $d\hat U_{\hbar}(\mathcal{E}(\mathcal{G}))$ of the enveloping algebra of the Lie algebra  $h_{3}$ generated by the operators $\{\hat Q_{j}\}_{j=1}^{3}$, $\{\hat P_{j}\}_{j=1}^{3}$ and $\hat I$. 

Let $\hat F_{j} = f_{j}(\hat Q_{1},\hat Q_{2},\hat Q_{3})$ be an $\scrS$-continuous operator in the O$^{\ast}$-algebra $\mathcal{A}$.   We can represent the operator $\hat F_{j} \in \mathcal{A}$ acting on $ \mathcal{S}(\RR^{3})$ by \begin{equation}\label{EQ74}
(f_{j}(\hat Q_{1},\hat Q_{2},\hat Q_{3})\psi)(\vec x) = f_{j}(\vec x) \psi(\vec x) \;\text{ for}\; \psi\in  \mathcal{S}(\RR^{3}),
\end{equation} because for any function $f_{j}(\vec x)$ that maps $\mathcal{S}(\RR^{3}) \to \mathcal{S}(\RR^{3})$, \begin{equation}
 \:\int |f_{j}(\vec x)\psi(\vec x)|^{2}d^{3}x < \infty\; \text{for all} \; \psi\in \mathcal{S}(\RR^{3}).
\end{equation}

The sequence of simultaneous approximate eigenfunctions $\{\Phi^{n}_{\vec \alpha}\}_{n=1}^{\infty},$ with $\Phi^{n}_{\vec \alpha} = \phi^{n}_{\alpha_{1}}\otimes \phi^{n}_{\alpha_{2}}\otimes \phi^{n}_{\alpha_{3}}$, for $\vec \alpha \in \RR^{3}$ in the continuous spectra of the commuting operators $(\hat Q_{1},\hat Q_{2},\hat Q_{3})$, determines a sequence of approximate eigenstates $\{\hat P_{\Phi^{n}_{\vec \alpha}}\}_{n=1}^{\infty}$ with, for each $j$, $Tr \hat P_{\Phi^{n}_{\vec \alpha}}\hat Q_{j} \to \alpha_{j}$ as $n\to \infty.$ Furthermore for each operator $\hat F_{j}\in \mathcal{A}$, \begin{equation}
Tr \hat P_{\Phi^{n}_{\vec \alpha}}\hat F_{j} = \langle \Phi^{n}_{\vec \alpha}, f_{j}(\hat Q_{1},\hat Q_{2},\hat Q_{3})\Phi^{n}_{\vec \alpha}\rangle 
\end{equation} where
\begin{equation}
\langle \Phi^{n}_{\vec \alpha}, f_{j}(\hat Q_{1},\hat Q_{2},\hat Q_{3})\Phi^{n}_{\vec \alpha}\rangle  = \int | \Phi^{n}_{\vec \alpha}(\vec x)|^{2} f_{j}(\vec x) d^{3}x
\end{equation}

\subsubsection{Examples}\label{examp} Examples of sequences of approximate eigenfunctions for the position operators $\hat Q_{j}$ include \begin{enumerate}\item those constructed using "cap-functions", defined by   \begin{itemize}
\item $\chi^{\sigma}(\vec x) =  \exp(-\sigma^{2}/(\sigma^{2}-||\vec x||^{2})),\;\;\text{if} \;\;||\vec x||< \sigma$, and 
\item $\chi^{\sigma} (\vec x) = 0,\; \text{if}\;\; ||\vec x||\geq \sigma.$  
\end{itemize} If $\int \chi_{\sigma} = c >0$, then $\frac{1}{c} \chi_{\sigma}$ is normalised, $\frac{1}{c}\int  \chi_{\sigma} = 1$, that is, \begin{equation}
\frac{\sigma^{3}}{c} \int _{B(\vec 0, 1)}\exp(- 1/(1 -\|\vec x\|^{2}))d^{3}x  = 1
\end{equation}
where $B(\vec 0, 1)$ is the open sphere centre at $\vec 0$, radius $1$.

The function $\frac{1}{c}\chi^{\sigma}_{\vec \alpha} (\vec x) = \frac{1}{c}\chi^{\sigma} (\vec x -\vec \alpha)$ has support in the open sphere $B(\vec \alpha, \sigma)$,  centred at $\vec \alpha \in \RR^{3}$ of radius $\sigma$. The graph of the cap-function reveals that as $\sigma \to 0$,  $\frac{1}{c}\chi^{\sigma}_{\vec \alpha}$ approaches $\delta(\vec x - \vec \alpha)$.

The functions $\psi^{\sigma}_{\vec \alpha}(\vec x) = ( \frac{1}{c}\chi^{\sigma}_{\vec \alpha}(\vec x))^{\frac{1}{2}}$ are normalised wave-functions of compact support. When $\sigma = \frac {1}{n}$ and $\vec \alpha = (\alpha_{1},\alpha_{2},\alpha_{3})$, the sequence of approximate eigen-functions $\{\Phi^{n}_{\vec \alpha}(\vec x) =\psi^{\frac{1}{n}}_{\vec \alpha}(\vec x) \}_{n=1}^{\infty}$ satisfies $\| \hat Q_{j}\Phi^{n}_{\vec \alpha} - \alpha_{j} \Phi^{n}_{\vec \alpha}\| \to 0$ as $n \to \infty$ for each $\alpha_{j} \in \sigma_{c}(\hat Q_{j})$. The corresponding sequence of eigenstates is given by the sequence of projection operators $\{\rho^{n}_{\vec \alpha} = \hat P_{\Phi^{n}_{\vec \alpha}}\}_{n=1}^{\infty}.$ 

Since the functions $\{f_{j}\}$ are continuous, \begin{equation}
Tr \hat P_{\Phi^{n}_{\vec \alpha}}\hat F_{j} = \frac{1}{c}\int \chi^{\frac{1}{n}}_{\vec \alpha} (\vec x)f_{j}(\vec x) d^{3}x \to f_{j}(\vec \alpha)
\end{equation} as $n \to \infty$. Given any $\epsilon > 0$ there exists a $\delta >0$ such that $|f_{j}(\vec x) - f_{j}(\vec \alpha) | < \epsilon$  when $\|\vec x - \vec \alpha \| < \delta$, but the cap-shaped function $ \frac{1}{c}\chi^{\frac{1}{n}}_{\vec \alpha} (\vec x)$ has support in the open sphere $B(\alpha, \frac {1}{n})$ and is normalised so that,  when $n > \frac{1}{\delta}$,  \begin{equation}
\frac{1}{c}\int \chi^{\frac{1}{n}}_{\vec \alpha} (\vec x)f_{j}(\vec x) d^{3}x - f_{j}(\alpha) = \frac{1}{c}\int \chi^{\frac{1}{n}}_{\vec \alpha} (\vec x) [f_{j}(\vec x) - f_{j}(\alpha)]d^{3}x < \epsilon.
\end{equation} Thus $ \lim_{n\to \infty}\hat P_{\Phi^{n}_{\vec \alpha}(\vec x)} = \delta(\vec x -\vec \alpha)$ as linear functionals on the commutative subalgebra of operators $\hat F_{j} \in \mathcal{A}$.

\item A {\em second example} of a sequence of approximate eigenfunctions is constructed from Gaussian wave functions, indexed by $n \in \NN$, for $j=1,2,3$,\begin{equation}
 \phi_{\alpha_{j}}^{n}(x_{j}) = (\frac{n^{2}}{2\pi})^{\frac{1}{4}} \exp \left \lbrack - n^{2}\frac{(x_{j} - \alpha_{j})^{2}}{4} \right \rbrack.
\end{equation} As the joint approximate eigenfunction 
$\Phi^{n}_{\vec \alpha}(\vec x) = \otimes _{j=1}^{3} \phi_{\alpha_{j}}^{n}(x_{j})$, its modulus squared is 
\begin{equation} |\Phi^{n}_{\vec \alpha}(\vec x)|^{2}= (\frac{n^{2}}{2\pi})^{\frac{3}{2}} \exp \left \lbrack - n^{2} \frac{\|\vec x- \vec \alpha\|^{2}}{2} \right \rbrack,
\end{equation} the graph of which approaches that of $\delta(\vec x - \vec \alpha)$ as $n \to \infty$. 

To show this, continuity of the polynomials is important but the Gaussian unlike the cap-function does not have compact support. However a Gaussian $|\Phi^{n}_{\vec \alpha}(\vec x)|^{2}$ is a normal probability distribution with mean $\vec \alpha$ and standard deviation $\frac{1}{n}$.
Chebyshev's inequality implies that the probability that $\|\vec x - \vec \alpha\| > \frac {k}{n} $ for some $k>0$ is less than $\frac{1}{k^{2}}$. Therefore as $n\to \infty$ the difference $\|\vec x - \vec \alpha\| $ remains bounded, 

For any function $f_{j}(\vec x)$, $\lim _{n\to \infty} \int d^{3}x  |\Phi^{n}_{\vec \alpha}(\vec x)|^{2} f_{j}(\vec x) = f_{j}(\vec \alpha)$ for by putting $\vec y = n(\vec x - \vec \alpha)$, $|\int d^{3}x  |\Phi^{n}_{\vec \alpha}(\vec x)|^{2} f_{j}(\vec x) - p(\vec \alpha)|$ becomes  
\begin{equation}
 |\int d^{3}y({2\pi})^{-\frac{3}{2}}\exp\frac{- \|\vec y\|^{2}}{2}[f_{j}(\vec \alpha + \frac{\vec y}{n}) - f_{j}(\vec \alpha) ] |
  \end{equation} so by continuity of the $f_{j}$, given $\epsilon > 0$ there exists a $\delta >0$ such that  $|f_{j}(\vec \alpha + \frac{\vec y}{n}) - f_{j}(\vec \alpha) | < \epsilon $ for $n > \frac{\|\vec y\|}{\delta}$.  Therefore as linear functionals on the commutative subalgebra of operators $\hat F_{j}\in \mathcal{A}$, in the $\lim n\to \infty$, \begin{equation}
\int d^{3}x  |\Phi^{n}_{\vec \alpha}(\vec x)|^{2} f_{j}(\vec x) \to f_{j}(\vec \alpha).
\end{equation} This example is the prototype of a family of Gaussian wave functions, labelled by $\vec \alpha$ and $\vec \beta \in \RR^{3}$, that are approximate position eigenstates, 
 \begin{equation} 
\Phi_{\vec \alpha, \vec\beta}^{n}(\vec x) = (\frac{n^{2}}{2\pi})^{\frac{3}{4}}\exp \left \lbrack - n^{2} \frac{\|\vec x- \vec \alpha\|^{2}}{4} \right \rbrack \exp (\imath \frac{\vec x \cdot \vec \beta}{\hbar})
\end{equation} all of whose modulus squared is  \begin{equation}
|\Phi_{\vec \alpha, \vec\beta}^{n}(\vec x)|^{2} =  |\Phi^{n}_{\vec \alpha}(\vec x)|^{2}
\end{equation} so that, in the Schr\"odinger representation, all converge as linear functionals to $\delta(\vec x - \vec \alpha).$
\end{enumerate}

\subsection{ The proof of equation(\ref{EQ100}) } \label{ocover} 
For any $\rho_{0} \in \mathcal{E}_{S}(\mathcal{A})$ and any $\epsilon > 0$ \begin{equation}
|Tr \rho \hat F_{l} - f_{l}(Tr \rho\hat Q_{1},Tr\rho \hat Q_{2},Tr \rho \hat Q_{3})| < \epsilon, \;\text{for}\; l = 1,2,3
\end{equation} for all $\rho \in  \nu(\rho_{0}, \delta _{0})$ provided that \begin{equation}\label{EQ121}
\delta_{0} = \min\lbrack \lbrace\frac{\delta_{j}}{\kappa_{j}(\hat Q_{j})}\rbrace_{j=1}^{3} \lbrace \frac{\epsilon}{3\kappa_{m_{l}}(\hat F_{l})}\rbrace_{l=1}^{3}\rbrack
\end{equation} in which, for each $j =1,2,3$, $\delta_{j} > 0$ is such that $|f_{l}(Tr \rho \hat Q_{j}) -f_{l}( \alpha_{j})| < \frac{\epsilon}{6}$ when $|Tr \rho \hat Q_{j} - \alpha_{j}| < \delta_{j}$ and the $\kappa_{m}(\hat A)$ are defined as in equation (\ref{EQ110}) and are calculated in \S \ref{relb} with $\kappa_{m}(\hat Q_{j}^{m}) = \frac{1}{2}$.

If $\alpha_{j} = Tr \rho_{0}\hat Q_{j}$ and $\epsilon > 0$ there are integers $N_{0}(\delta_{j} )$ and $N_{1}(\epsilon )$, such that $ n >  N_{0}(\delta_{j} )$ implies that $| Tr \rho^{n}_{\vec \alpha} \hat Q_{j} - \alpha_{j}| < \delta_{j},$ $j = 1,2,3$, see equation (\ref{EQ43}), where the $\delta_{j} >0$ are such that, for $l=1,2,3$, \begin{equation}\label{EQ15}
|f_{l}(Tr \rho \hat Q_{1}, Tr \rho \hat Q_{2},Tr \rho \hat Q_{3}) - f_{l} (Tr \rho_{0} \hat Q_{1}, Tr \rho_{0}  \hat Q_{2},Tr \rho_{0}  \hat Q_{3}) | < \frac{\epsilon}{3}.
\end{equation}  and $ n >  N_{1}(\epsilon )$ implies that, by equation (\ref{EQ44}),\begin{equation}\label{EQ153} | Tr \rho^{n}_{\vec \alpha} \hat F_{l} - f_{l}(\vec \alpha )| < \frac{\epsilon}{6}\;\; \text{for}\;\; l = 1,2,3.\end{equation}

If $\delta_{0} >0$ satisfies equation (\ref{EQ121}), the open set $ \nu(\rho_{0}, \delta _{0})$ 
\begin{enumerate} 
\item contains, for $n > \max( \{N_{0}(\delta_{j} )\}_{j=1}^{3}, N_{1}(\epsilon ))$,the approximate eigenstates  $\rho_{\vec \alpha}^{n} = \hat P_{\Phi^{n}_{\vec \alpha}} $ of the operators $\{\hat Q_{j}\}_{j=1}^{3}$, at $\{\alpha_{j} \in \sigma_{c}(\hat Q_{j})\}_{j=1}^{3}$.
\item is such that when $\rho \in  \nu(\rho_{0}, \delta _{0})$ then  \begin{equation}\label{EQ15}
|f_{l}(Tr \rho \hat Q_{1}, Tr \rho \hat Q_{2},Tr \rho \hat Q_{3}) - f_{l} (Tr \rho_{0} \hat Q_{1}, Tr \rho_{0}  \hat Q_{2},Tr \rho_{0}  \hat Q_{3}) | < \frac{\epsilon}{6}.
\end{equation} 
\end{enumerate}

\begin{theorem} On $V = \nu(\rho_{0}, \delta _{0})$, $F_{l}|_{V}$ is the locally linear approximation to $f_{j}(q_{1}|_{V}, q_{2}|_{V},q_{3}|_{V})$, that is, given $\epsilon > 0$ for all $\rho \in  \nu(\rho_{0}, \delta _{0})$ \begin{equation}\label{EQ54}
|Tr \rho \hat F_{l} - f_{l}(Tr \rho \hat Q_{1}, Tr \rho \hat Q_{2},Tr \rho \hat Q_{3}) | < \epsilon.
\end{equation} \end{theorem}
 \begin{proof}
Writing $Tr \rho \vec Q$ in place of $(Tr \rho \hat Q_{1}, Tr \rho \hat Q_{2},Tr \rho \hat Q_{3}) $, it follows that $| f_{l}(Tr \rho \vec Q) - Tr \rho \hat F_{l} | $ is less than
\begin{equation}
 |f_{l}(Tr \rho \vec Q) - f_{l} (Tr \rho_{\vec \alpha}^{n}  \vec Q)|+ |f_{l} (Tr \rho_{\vec \alpha}^{n}  \vec Q) - Tr \rho_{\vec \alpha}^{n}  \hat F_{l}| +  |Tr \rho_{\vec \alpha}^{n}  \hat F_{l} - Tr \rho \hat F_{l} |.
\end{equation} Now, by equation(\ref{EQ15}), for the continuous functions $f_{l}$ and $\delta_{0}$ satisfying equation (\ref{EQ121}),  for all  $\rho \in  \nu(\rho_{0}, \delta _{0})$ \begin{equation}
 |f_{l}(Tr \rho \vec Q) - f_{l} (Tr \rho_{\vec \alpha}^{n}  \vec Q)| < \frac{\epsilon}{3},
\end{equation} 
and with $n > N_{1}(\epsilon)$, by equation (\ref{EQ441}) and by equation(\ref{EQ45}), \begin{equation}
|f_{l} (Tr \rho_{\vec \alpha}^{n}  \vec Q) - Tr \rho_{\vec \alpha}^{n} \hat F_{l}| < |f_{l} (Tr \rho_{\vec \alpha}^{n}  \vec Q) - f_{l}(\vec \alpha)| + |f_{l}(\vec \alpha) - Tr \rho_{\vec \alpha}^{n}  \hat F_{l}|
\end{equation}where both $|f_{l} (Tr \rho_{\vec \alpha}^{n}  \vec Q) - f_{l}(\vec \alpha)| < \frac{\epsilon}{6}$ and $|f_{l}(\vec \alpha) - Tr\rho_{\vec \alpha}^{n}  \hat F_{l}| < \frac{\epsilon}{6}$ and for all $\rho \in  \nu(\rho_{0}, \delta _{0})$, using equation (\ref{EQ121}), for all $n > \max(\{N_{0}(\delta_{j} )\}_{j=1}^{3}, N_{1}(\epsilon))$,
\begin{equation}
 |Tr\rho_{\vec \alpha}^{n}  \hat F_{l} - Tr \rho \hat F_{l} | < \kappa_{m_{l}}(\hat F_{l})Tr|\rho_{\vec\alpha}^{n} - \rho| < \frac{\epsilon}{3}.
\end{equation}
Thus if $n > \max( \{N_{0}(\delta_{j} )\}_{j=1}^{3}, N_{1}(\epsilon ))$,
then $|Tr \rho \hat F_{l} - f_{l}(Tr \rho \vec Q) | < \epsilon$ for all $\rho \in  \nu(\rho_{0}, \delta _{0})$. \end{proof}

\begin{theorem}
The family of open sets $\nu(\rho_{0}, \delta _{0})$, for all $\rho_{0}\in \mathcal{E}_{S}(\mathcal{A})$,  form an open cover of $\mathcal{E}_{S}(\mathcal{A})$ when  $
\delta_{0} = \min\lbrack \lbrace\frac{\delta_{j}}{\kappa_{j}(\hat Q_{j})}\rbrace_{j=1}^{3} \lbrace \frac{\epsilon}{3\kappa_{m_{l}}(\hat F_{l})}\rbrace_{l=1}^{3}\rbrack$ where, for each $j =1,2,3$, $\delta_{j} > 0$ is such that $|f_{l}(Tr \rho \hat Q_{j}) -f_{l}( \alpha_{j})| < \frac{\epsilon}{6}$ when $|Tr \rho \hat Q_{j} - \alpha_{j}| < \delta_{j}$ and the semi-norms $\kappa_{m}(\hat A)$ are defined as in equation (\ref{EQ110}).\end{theorem}
\begin{proof} Since the family contains every state, pure or mixed, labelled as $\rho_{0}$, we have only to check that $\delta_{0} > 0$. This follows because each of the six real numbers $\lbrack \lbrace\frac{\delta_{j}}{\kappa_{j}(\hat Q_{j})}\rbrace_{j=1}^{3} \lbrace \frac{\epsilon}{3\kappa_{m_{l}}(\hat F_{l})}\rbrace_{l=1}^{3}\rbrack$ is greater than zero.

\end{proof}
 
 Therefore the infinitesimal qr-number equations of motion for a massive particle are 
\begin{equation}\label{EQ349}
 \frac{d}{dt}Tr \rho\hat Q_{j} = \frac{1}{m}Tr \rho \hat P_{j}\; \text{and}\; \frac{d}{dt}Tr \rho \hat P_{j} = Tr \rho \hat F_{j}
\end{equation} where the operator $\hat F_{j} = f_{j}(\hat Q_{1},\hat Q_{2},\hat Q_{3}).$ Since the equations hold for all $\rho \in \mathcal{E}_{S}(\mathcal{A})$, Heisenberg's operator equation follow and conversely.

\subsection{ A proof of relative boundedness}\label{relb}
Each of the operators $\hat Q_{j}$ is relatively bounded, Kato \cite{14}, \S V.4.1,  with respect to the operator $\hat L = \sum_{j=1}^{3}(\hat P_{j}^{2} + \hat Q_{j}^{2}) + \hat I$ with relative bound $\frac{1}{2}$ so that $\kappa_{1}(\hat Q_{j}) = \frac{1}{2}$ for each $j=1,2,3$.
\begin{proof}
Since $(x -1)^{2} \geq 0$ for $x\in \RR$ then $x^{2} \leq \frac{1}{2}x^{4} + \frac{1}{2} $ so that if $x^{2} =\langle \phi , \hat L  \phi\rangle$  for all unit vectors $\phi \in \mathcal{D}^{\infty}(\hat U)$,
\begin{equation} \langle \phi, \hat Q_{k}^{2} \phi\rangle \leq  \langle \phi,\hat L \phi\rangle \leq  \frac{1}{2}\langle \phi , \hat L \phi\rangle^{2} + \frac{1}{2} \|\phi\|^{2}
\end{equation} whence, as $\langle \phi, \hat L\phi\rangle ^{2} \leq \| \hat L\phi\|^{2}$,\begin{equation}
\|\hat Q_{k}\phi\|^{2} \leq \frac{1}{2}\| \hat L \phi\|^{2}  +  \frac{1}{2}\|\phi\|^{2}
\end{equation} From this it is easily to deduce that  
\begin{equation}
\|\hat Q_{k}\phi\| \leq \frac{1}{2}\| \hat L \phi \|  +  \frac{1}{2}\|\phi\| = \frac{1}{2}\| (\hat L  - \imath\hat I)\phi \|.
\end{equation} Thus equation (\ref{EQ110}), with $d\hat U((I -\Delta))^{-1}$,  representing  the resolvent operator of  $\hat L $, gives  $\kappa_{1}(\hat Q_{j}) = \frac{1}{2}$ for each $j=1,2,3$.
\end{proof} The same argument works for the momentum operators $\kappa_{1}(\hat P_{j}) = \frac{1}{2}$, for $j=1,2,3$.

A similar argument will work for the tensor product, $\otimes _{j=1}^{3}\hat Q_{j}$, because $ \langle \Phi,\otimes _{j=1}^{3}\hat Q_{j} \Phi \rangle\leq  \langle \phi,\hat L \phi\rangle \leq  \frac{1}{2}\langle \phi , \hat L \phi\rangle^{2} + \frac{1}{2} \|\phi\|^{2}$ for all unit vectors  $\Phi \in \mathcal{D}^{\infty}(\hat U)$. The argument yields $\kappa_{(1,1,1)}(\otimes _{j=1}^{3}\hat Q_{j}) = \frac{1}{2}$ and for $\otimes _{j=1}^{3}\hat Q_{j}^{m}$, $\forall m \in \NN$, taking $x^{2} = \langle \phi , \hat L^{m}  \phi\rangle$  gives $\kappa_{(m,m,m)}(\otimes _{j=1}^{3}\hat Q_{j}^{m}) = \frac{1}{2}$.



\end{document}